\documentclass[graphicx,11pt]{aastex}

\lefthead{}
\righthead{}

\begin{document}

\title{NOTE:  Explaining why the Uranian satellites have equatorial prograde
orbits despite the large planetary obliquity.} 

\author{\textbf{A. Morbidelli$^{(1)}$, K. Tsiganis$^{(2)}$,
    K. Batygin$^{(3)}$, A. Crida$^{(1)}$ and R. Gomes$^{(4)}$ }\\  
(1) Laboratoire Lagrange, UMR7293, Universit\'e de Nice Sophia-Antipolis,
  CNRS, Observatoire de la C\^ote d'Azur. Boulevard de l'Observatoire,
  06304 Nice Cedex 4, France. (Email: morby@oca.eu / Fax:
  +33-4-92003118) \\
(2) Department of Physics, Section Astrophysics, Astronomy and Mechanics;
Aristotle University of Thessaloniki. Thessaloniki, GR 54124.
(E-mail : tsiganis@auth.gr / Fax: +30-31-995384) \\ 
(3) California Institute of Technology, Division of Geological \& Planetary Sciences;
MC 170-25 1200 E. California Blvd., Pasadena, CA 91125. (Email: kbatygin@gps.caltech.edu) \\ 
(4) Observat\'orio Nacional Jos\'e Cristino , Rua General José Cristino 77, CEP 20921-400, Rio de Janeiro, RJ, Brazil. (Email: rodney@ob.br)} 

\begin{abstract}
We show that the existence of { prograde} equatorial satellites is
{ consistent} with { a} collisional tilting scenario for Uranus.
In fact, if the planet was surrounded by a proto-satellite disk at the
time of the tilting and a massive ring of material was temporarily
placed inside the Roche radius of the planet by the collision, the
proto-satellite disk would have started to precess incoherently around
the equator of the planet{, up to a distance greater than that of
  Oberon}.  Collisional damping would then have collapsed it into a
thin equatorial disk, from which the satellites eventually formed. The
fact that the { orbits of the} satellites are prograde requires
Uranus to have had a non-negligible initial obliquity (comparable to
that of Neptune) { before} it was finally tilted to 98 degrees.
\end{abstract}

\section{Introduction}

The origin of the large obliquity of Uranus remains elusive. Two
scenarios have been proposed: an impulsive tilt due to a collision
with a massive body (Safronov, 1966) or a slow tilt due to a resonance
between the precession rates of the spin axis and of the orbit (Bou\'e
and Laskar, 2010).

A critical constraint, inherent to both of these scenarios, is that
the regular satellites of Uranus have essentially equatorial orbits
and are prograde relative to the rotation of the planet. Notice that
the rotation of the planet is, strictly speaking, retrograde, as
Uranus obliquity is about 98 degrees.

In principle, if the satellites were originally coplanar to the
equator of Uranus and the planet was tilted slowly (as in Bou\'e and
Laskar, 2010), the satellites would have preserved equatorial orbits
by adiabatic invariance. Indeed, in the system's current
configuration, the Laplace plane ({the} reference plane about which
satellite orbits precess) is very close to Uranus' equatorial plane,
for all bodies up to Oberon's distance, due to the oblateness of the
planet.

In order to tilt Uranus slowly, a resonance between the precession
rates of the spin axis and of the orbital plane is required. This
means that the former had to be much faster than it is today. In
Bou\'e and Laskar (2010) this is achieved by assuming that Uranus
originally had a massive satellite with an orbital radius of about
0.01 AU (Satellite X, hereafter). This assumption, however, is
problematic because the Laplace plane for Satellite X is close to the
orbital plane of Uranus. Thus, Satellite X would not follow the
equator during the tilting of the planet and, by virtue of its large
mass, would retain the other satellites (particularly Titania and
Oberon, as we verified by numerical integration of a slowly tilting
system) { near} its own orbital plane.  When Satellite X is removed by
chaotic dynamics, the tilting process is over. Yet, the orbits of the
regular satellites of Uranus would remain off equator, as in the
impulsive tilting scenario.

In absence of a slow-tilting scenario that does not invoke the
existence of Satellite X, we are left with the impulsive tilting
scenario as the only viable option. Accordingly, in this Note, we
investigate the conditions under which the equatorial, prograde orbits
of the regular satellites of Uranus can be reproduced in the context
of the collisional tilting scenario. 

{For clarity, we proceed in steps.  We first investigate in
  Sections~2\&3 the dynamics of a proto-satellite disk around an
  oblated planet with a tilted spin axis. This highlights the
  competing effects of the planet's $J_2$, the solar perturbation and
  the self-gravity of the disk. We focus on the conditions that lead
  the disk to precess incoherently around the planet's equatorial
  plane and, eventually, to collapse into an equatorial disk. We don't
  worry at this stage on whether this equatorial disk would be
  prograde or retrograde. Thus, although we fix the obliquity at the
  current value for Uranus, the dynamics that we study are basically
  independent on the obliquity value. Then, in Section~4 we consider
  the fact that Uranus is a retrograde planet (obliquity
  $\epsilon=98^\circ$). Thus, a disk originally on the orbital plane
  of Uranus (as expected if Uranus was tilted from $\epsilon=0^\circ$
  to $98^\circ$ in one shot) would necessarily become an equatorial,
  {\it retrograde} disk. We then investigate which tilting histories
  of the planet would have a non-negligible probabilities to produce a
  {\it prograde} equatorial disk. In Section 5, we finally discuss the
  implications of these tilting histories on our understanding of
  giant planets growth.}

\section{Dynamics of a proto-satellite disk around a tilted planet}

It is likely that the giant collision that tilted Uranus occurred
during the accretion phase of the planet, when satellites were not yet
formed and the planet was surrounded by a tenuous disk of gas, very
rich in solids, similar to that usually invoked for the formation of
satellites around giant planets (Canup and Ward, 2002). Moreover, even if the
planet already had a system of regular satellites at the time of
tilting, it is likely that the said system would have become
unstable\footnote{We checked this with simple $N$-body simulations,
  which assumed {Uranus tilted by 98 degrees and surrounded by 
  a system of regular satellites with the current masses and semi
  major axes within 1\% of the current ones, but orbits co-planar with
  Uranus' orbital plane} (hereafter denoted for brevity as ``the plane
  of the Sun'').}, developing mutually-crossing orbits. Presumably, the
satellites would have then collided with each other, generating a
debris disk.

For these reasons, we conduct our investigation assuming for
simplicity that, at the time of tilting, the planet was surrounded by
a proto-satellite disk of planetesimals.  Our nominal disk extends to
1.5 { times} the distance of Oberon, (i.e. up to $6\times 10^{-3}$ AU), and has a
mass equal to the combined masses of the five main current regular
satellites, (i.e. $10^{-4}$ Uranus masses, $M_U$). The surface density
profile of the disk is assumed to be inversely proportional to the
distance from the planet.  We assume that, initially, the spin of
Uranus is orthogonal to the plane of the Sun and that the disk lays on
such a plane.

The dynamical response of such a proto-satellite disk to the impulsive
tilting of the planet is far from trivial.  Here, we have utilized a
simple Laplace-Lagrange-like secular model, where we partition the
disk into a system of $N = 100$ { axisymmetric} massive rings that
interact with each other gravitationally (see Chapter 7 in Murray and
Dermott, 1999). The diagonal terms of the $A$ and $B$ matrices
entering the equations (see eq. (7.15) in Murray and Dermott, 1999)
are modified to account for the $J_2$ term of the potential of the
tilted planet, i.e. (see eq. (6.255) in Murray and Dermott, 1999):
\begin{equation}
A_{j,j}=A_{j,j} + n \frac{3}{2} J_2 \left(\frac{R_p}{a}\right)^2\cos\epsilon\ ,
\quad
B_{j,j}=B_{j,j} - n \frac{3}{2} J_2 \left(\frac{R_p}{a}\right)^2\cos\epsilon\ ,
\label{our-coef}
\end{equation}
where $R_p$ is the radius of Uranus, $a$ is the semi major axis of
the ring $j$, and $n$ is its orbital frequency. In the formula above,
the usual terms associated to $J_2$ { are} multiplied by
$\cos\epsilon$, where $\epsilon$ is the obliquity of { Uranus' spin axis}, to
account for the weakened effect of the oblateness of the planet for 
large obliquities. { We also took into account the scale-height, $h$, of the disk, by 
using properly softened Laplace coefficients, as in Hahn (2003). The results 
described below are practically the same for disks with $h=0.01-0.05$}

The situation where the planet is suddenly tilted, but the disk
initially still lays on the plane of the Sun, is modeled by assuming
that the rings initially have  $I=\epsilon$ and  $\Omega=0^\circ$,
where $I$ and $\Omega$ are the inclination and longitude of node
measured with respect to the equatorial plane of the planet. 
In addition, all rings are assumed to be
initially circular. The Sun is modeled with an additional ring,
at a distance of 20~AU, also with $I=\epsilon$ and $\Omega=0^\circ$.
The system is then evolved, according to the Laplace-Lagrange secular
equations. 

\begin{figure}[t!]
\centerline{\includegraphics[height=5.cm]{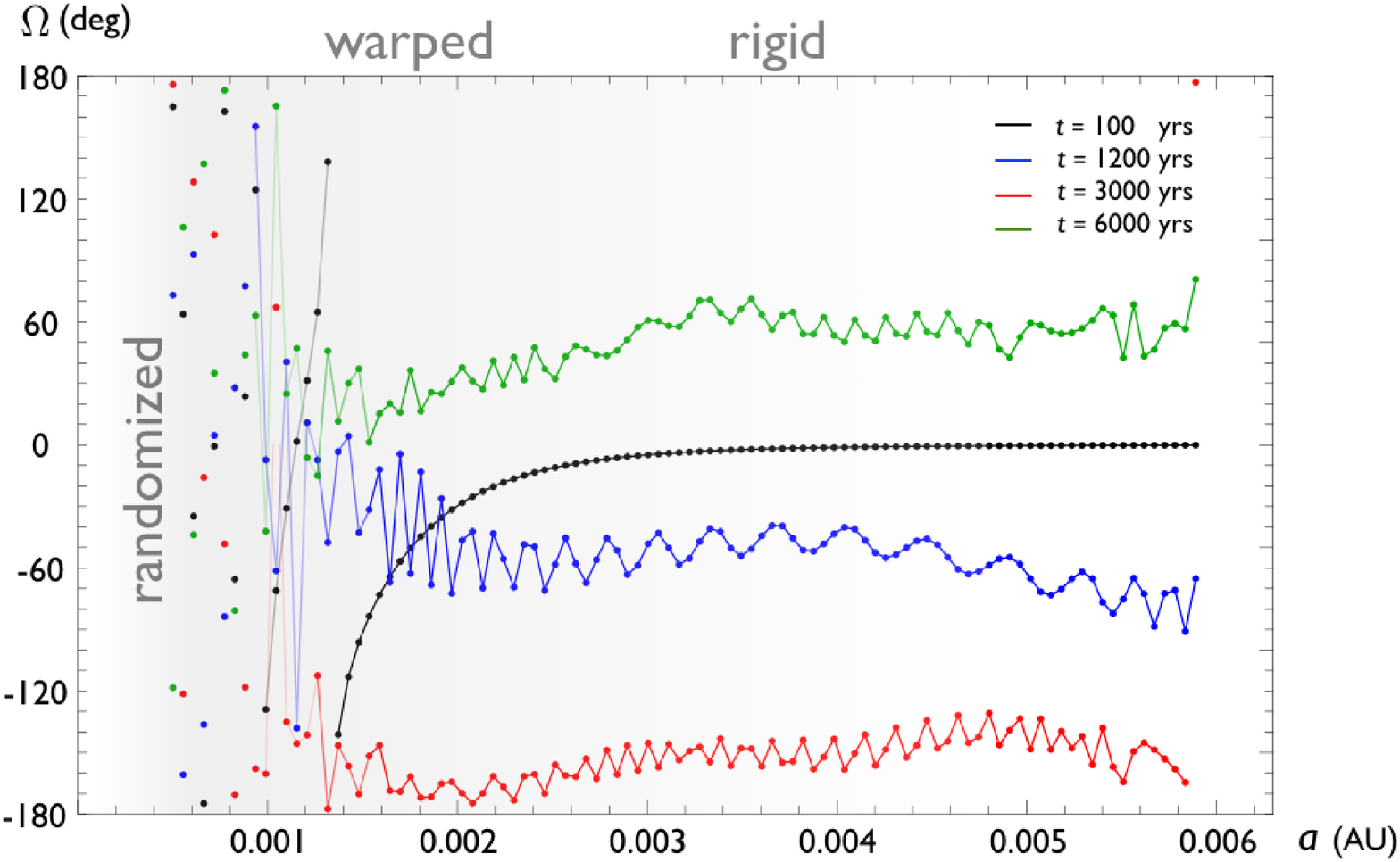}\quad
  \includegraphics[height=5.cm]{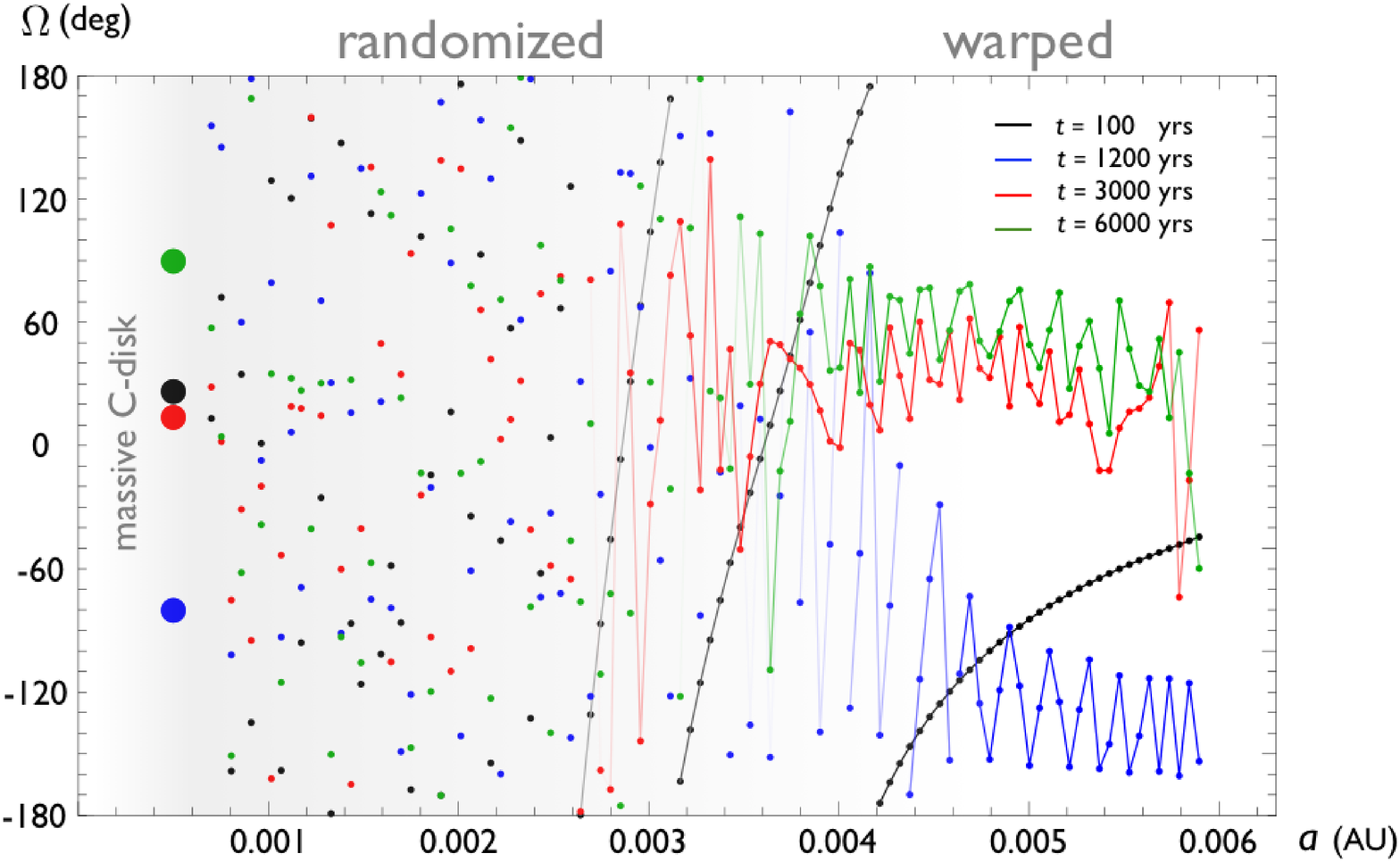}}
\caption{\small Longitudes of the nodes of the concentric rings that
  compose the proto-satellite disk in our Lagrange-Laplace model, as a
  function of the the rings' distance to the planet. Diffent colors
  represent different times, as labelled. The regions where the nodes
  are randomized, the disk is warped or precesses quasi-rigidly are
  also labelled. Left panel: the nominal case, with Uranus current
  $J_2$. Right panel: the case where a transient equatorial
  collisional disk of 0.01~$M_U$ is present at $\sim 3$ Uranian
  radii. The node of this disk is represented by a filled dot.}
\label{analytic}
\end{figure}

The left panel of Fig.\ref{analytic} shows the longitude of the node
of each ring, as a function of the ring's distance to Uranus, at four
different epochs represented by different colors. {The obliquity of Uranus $\epsilon$ is assumed to be $98^\circ$.} In the inner part of
the disk the longitudes of the node are rapidly randomized. This means
that the disk loses its disk-like structure and forms a thick torus
around the planet. The planetesimals in this portion of the disk can
then collide. Collisional break-ups and/or inelastic bouncing would
eventually damp the orbital inclination relative to the symmetry plane
of the dynamics, i.e. the equatorial plane of the planet. This would
form a thin equatorial disk, from which equatorial satellites can
eventually accrete. However, Fig.\ref{analytic} shows that this
intuitive scenario works only up to a distance of 0.001~AU from the
planet, i.e. slightly beyond the orbit of Miranda. Beyond this
threshold, the longitudes of the nodes of the rings are not
randomized. { The rings} precess, but their { nodal} values at any
given time trace a relatively smooth function of the distance from the
planet, with some small-amplitude oscillations.  This means that the
disk, under the effect of its own self-gravity, is preserving its
disk-like structure, precessing coherently around the planet's
equator.
	
More precisely, close to the 0.001~AU threshold, the disk is
significantly warped (i.e. the values of $\Omega$ change significantly
with the distance from Uranus), but beyond $\sim 0.003$~AU the disk is
characterized by almost rigid precession (i.e. all values of $\Omega$
are about the same). The inclinations (not shown in the plot) of the
rings in the region where the disk precesses rigidly remain
practically constant over the precession of $\Omega$. However, in the
warped region, they show regular oscillations with non-negligible
amplitude, related to the precession of $\Omega$. Thus, at any given
time, the inclinations change smoothly with $\Omega$ and the distance
from the planet. 

{Two caveats about this calculation needs to be discussed. First,}
the validity of the Laplace-Lagrange equations for this problem may
appear doubtful, because the inclinations $I$ of the rings are large,
whereas the equations have been developed asymptotically for systems
with $e,I$ that tend to 0. In reality, the inclinations are large {\it
  relative to the planet's equator}, and this fact is taken into
account by multiplying the coefficients in (\ref{our-coef}) by
$\cos\epsilon$. But the relative inclinations of the rings are
originally very small, in accordance with the Laplace-Lagrange theory.
Thus, the equations are valid and will remain valid as long as the
relative inclinations of the rings remain small, which is for all
times in the portion of the disk that precesses coherently, beyond
0.001~AU. The Lagrange-Laplace approximation breaks down in the inner
part of the disk, once the nodes get randomized. However, this is
enough for our purpose, which is just to investigate whether node
randomization takes place and where.

{The second caveat concerns our assumption of an initial circular
  and coplanar disk. In fact, Parisi and Brunini (1997) showed that
  the impulsive tilting of Uranus implies an impulsive change in the
  orbital velocity of the planet of about 2km/s. As a result,
  satellites or disk particles originally beyond 70-90 Uranus radii
  (1.2--1.5$\times 10^{-2}$ AU) would be removed from the system. The
  disk that we consider is much smaller than this
  threshold. Nevertheless, disk particles would get eccentric and/or
  inclined relative to the new orbital plane of the planet, the
  magnitude of $e$ and $i$ growing as $\sqrt{a}$. However, the
  trajectories of the disk particles would start to intersect with
  each other. This would lead to a rapid collisional damping, until
  the disk recovers its circular and coplanar structure. For this
  reason, we think that our initial conditions (a circular and
  coplanar disk) are nevertheless appropriate for our goal,
  i.e. determining the distance from the planet within which the
  orbital planes of the disk particles are randomized.}
 
The conclusion that we draw from Fig.~\ref{analytic} is that the
precession of the disk relative to the equator alone cannot explain
the equatorial orbits of the satellites beyond Miranda. In fact, a
disk precessing coherently, as that illustrated in the left panel of
Fig.\ref{analytic} beyond 0.001~AU, would eventually form satellites
on {\it its own plane}, i.e. on highly-inclined orbits relative to the
equator of the planet.

Note that our nominal disk represents a best-case scenario for the
randomization of the nodes. Had we chosen a disk less radially
extended (up to $4\times 10^{-3}$~AU, for instance) or more massive,
the effect of self-gravity would have been even more important and the
region where the nodes randomize would have been confined closer to
the planet. This strengthens the conclusion presented above. This
implies that the equatorial configuration of the Uranian satellites
requires a more complex explanation.

\section{An enhanced $J_2$ for the just-tilted Uranus}

The results from the previous section highlight the need for a
dynamical mechanism that can {\it broaden} the region where the nodes
of the disk particles get randomized, {because only in this region the
  disk loses coherence and can collapse on the planet's equatorial
  plane}. Recall that the disk loses coherence at a distance from the
planet where the precession rate around the equator forced by the
$J_2$ term dominates the precession rate around the disk's mid-plane {
  that is} forced by the disk's self-gravity. Because we cannot reduce
the self-gravity of the proto-satellite disk, the only option is to
find a mechanism that enhances Uranus' $J_2$ in the aftermath of the
collision.

A first idea that we explored is that the inner part of the disk,
within 0.001~AU damps on the equator and forms an equatorial satellite
({with a mass comparable to that of} Miranda). This enhances the
effective $J_2$ of the planet felt by the outer portion of the disk, {
  thus extending the zone in which the disk precesses incoherently,
  leading to the collapse of another ring on the equator, etc.}
However, { we found that this mass} is too small, and this effect is
negligible.

We then turned to published simulations of the collisional tilting {
  scenario for} Uranus. These simulations (Slattery et al., 1992) show
that the impact should have generated an equatorial disk of debris,
accounting for $\approx$1 to 3\% of the mass of Uranus, namely about
100 times more massive than the proto-satellite disk, but mostly
confined within 3 Uranian radii. We call this disk the C-disk,
equivalent to the proto-Lunar disk, as it was generated in the
collision, in order to avoid confusion with the proto-satellite disk
considered up to now.

The C-disk could {\it not} generate the current regular satellites of
Uranus, because the latter are too far away (Canup and Ward, 2000). In fact,
most likely as the C-disk spread outside of the Roche lobe of the
planet, it formed satellites (Charnoz et al., 2010) which, being situated
inside the corotation radius, tidally migrated onto the planet. The
existence of { this massive} C-disk (or of the close satellites that it
generated), however, is equivalent to increasing the planet's $J_2$
enormously. 

The right panel of Fig.\ref{analytic} shows the result of our
Laplace-Lagrange model of the evolution of the proto-satellite disk,
when we account for a C-disk of 0.01~$M_U$. {Again, the obliquity of Uranus is $\epsilon=98^\circ$.} The C-disk is represented
by a circular ring at 3 Uranian radii with $I=0$, but with a mass
multiplied by a factor $\cos\epsilon$ to account for its inclination
$\epsilon$ relative to the proto-satellite disk, as we did for the
$J_2$ coefficients in (\ref{our-coef}). As one can see from the plot,
the region where the longitude of nodes in the proto-satellite disk
are randomized now extends to 0.003~AU (almost the distance of
Oberon). The radial extent of the randomized disk can be enhanced
further by assuming a more massive C-disk. For example, if we take the
mass of the C-disk to be 0.03~$M_U$ the randomized region extends to
0.0045~AU and so on.

We have checked our results with N-body simulations that take the
self-gravity of the proto-satellite disk into account. The C-disk is
modeled with a single massive satellite on a circular orbit at 3
Uranian radii on the equatorial plane of the planet. The
proto-satellite disk is modeled with { 2,000-8,000} equal-mass
particles, initially on circular orbits on the plane of the Sun. All
particles interact with each other, with a smoothing length equal to {
  0.5-3 Hill spheres of the innermost particle, which is a equivalent
  to a few times the particle radius. We also tested disks twice as
  massive as our nominal case and with different initial scale
  heights. We performed several simulations and found our results to
  be robust with respect to changes in the values of the above
  parameters. The calculations were performed using a version of SyMBA
  (Duncan et al., 1998), suitably modified so that the $N^2$ force
  calculation is performed on a GPU card, with the help of the SAPPORO
  library (Gaburov et al., 2009). The time-step was chosen equal to 1/30 times
  the period of the massive satellite.}

\begin{figure}[t!]
\centerline{\includegraphics[height=5.cm]{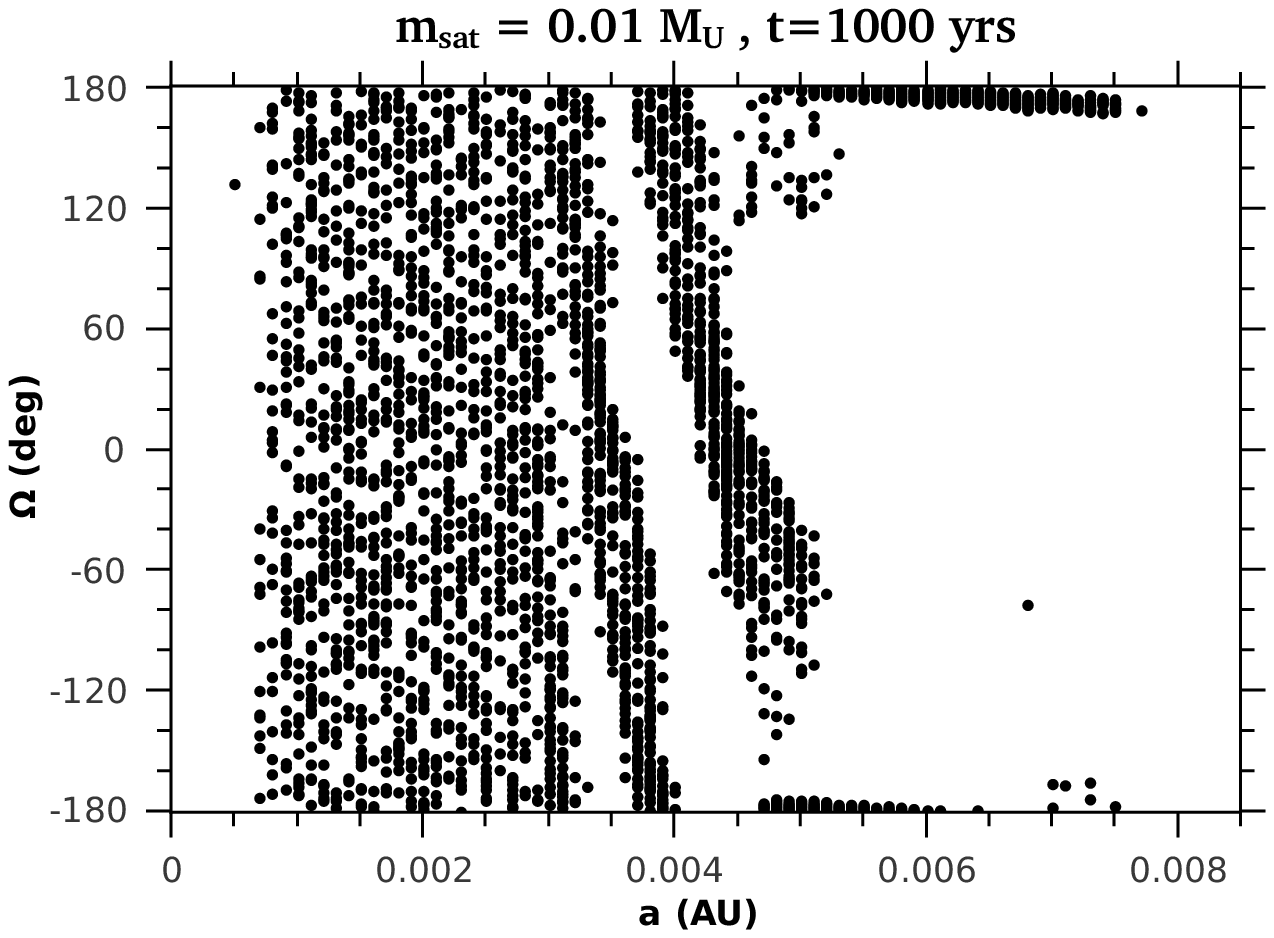}\quad
  \includegraphics[height=5.cm]{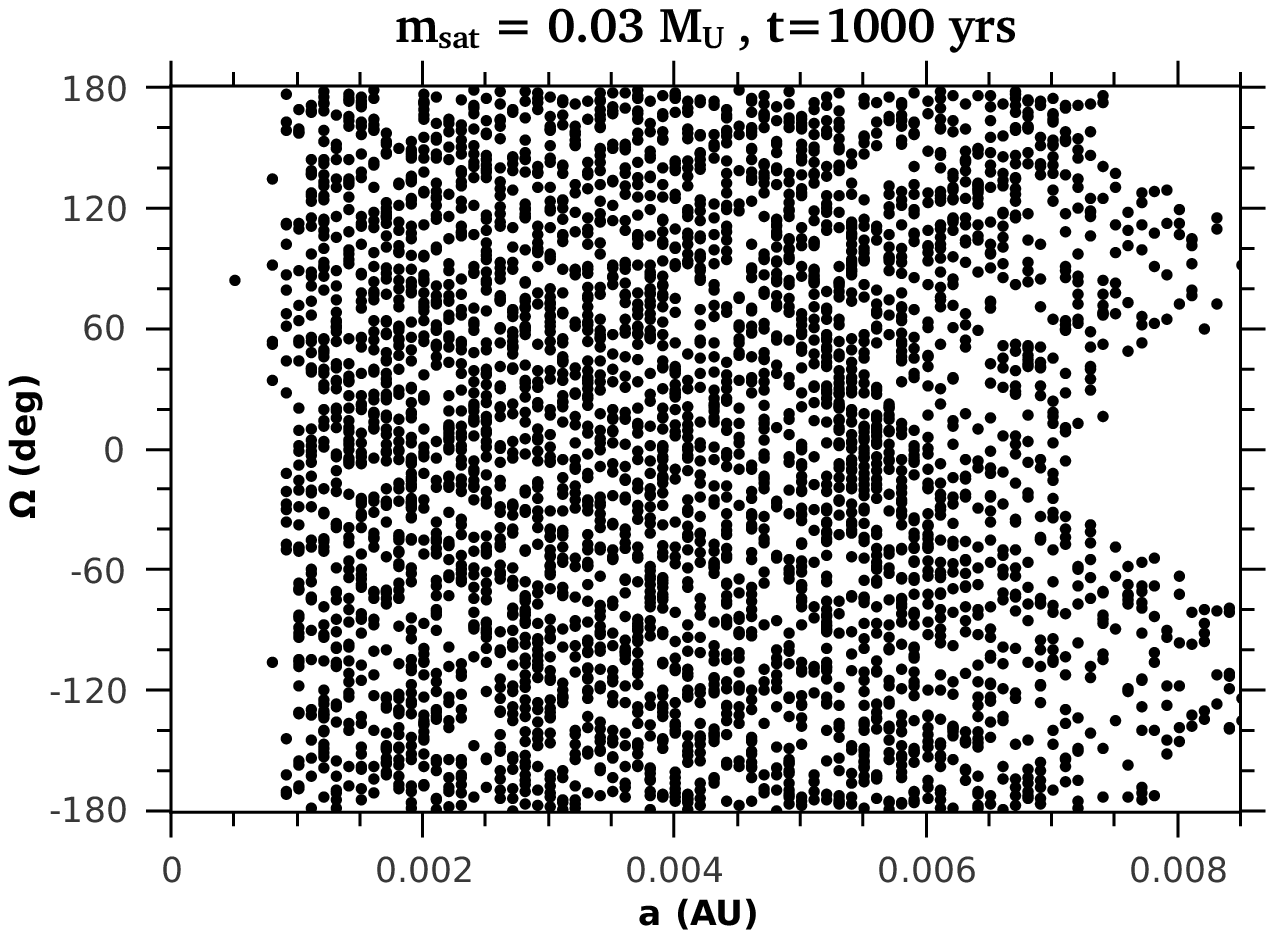}}
\caption{\small Longitudes of the nodes of the disk particles as a
  function of their semi major axis, after 1000y years of numerical
  simulation. The left panel includes the effect of a 0.01~$M_U$
  satellite at 3 Uranian radii. The right panel is for a 0.03~$M_U$
  satellite.}
\label{numeric}
\end{figure}

The results { of two simulations (4,000 particles, nominal mass) are } shown 
in Fig.~\ref{numeric}. In the case where the
C-disk has a mass of 0.01~$M_U$ (left panel), after a time of $1000~$yr, the randomized 
region extends to 0.003~AU, in excellent agreement with our analytic model. Beyond 
this threshold, though, the disk is extremely warped, up to about 0.005~AU
(i.e. beyond the current position of Oberon). By mutual interactions,
the particles in the 0.003-0.005~AU region also become quite eccentric
and cross each other { (the Laplace-Lagrange approximation is not valid there)}. 
On the longer run, therefore, the randomization
is expected to extend to 0.005~AU.  Instead, in the case where the
C-disk has a mass of 0.03~$M_U$ (right panel), the { nodes of the disk particles are} randomized up
to 0.006~AU within the { same} timescale. 

We conclude from these experiment that, if the collision that tilted
Uranus really formed a massive, equatorial transient disk, as in the
simulations of Slattery et al. (1992), the proto-satellite disk would have lost
coherence up to { a distance at least equal to that of Oberon},
dispersing symmetrically around the equator of the planet. Collisional
damping would have then {forced the proto-satellite disk to collapse}
onto the equatorial plane, where the observed satellites, from Miranda
to Oberon, could form. { It is likely that, during this process, some
  of the disk's mass would be ``lost'', instead of accreting on the
  satellites. However, our simulations give practically the same
  results even for a disk twice as massive as the combined mass of the
  current satellite system. This is because the mass ratio between the
  C-disk and the proto-satellite disk is of $\sim 100$.}

\section{Why are Uranus satellites prograde?}

If Uranus had been tilted abruptly from an obliquity of 0 to one of 98
degrees, the mechanism described in the previous section would have
produced a system of equatorial, but retrograde satellites. This is because
the precession of the particles of the proto-satellite disk around the equator 
and their collisional damping 
would preserve the total angular momentum of the disk relative to the spin
axis of the planet, which is negative from the very beginning (i.e.\ at
the time when the planet is tilted and the disk is still on the plane
of the Sun). 

To have the disk collapse by collisional damping onto an equatorial,
{\it prograde} disk, it is necessary that the obliquity of Uranus was
not null when the planet's { final tilting episode} occurred. Let's
call $\epsilon'$ this pre-final-tilt obliquity.

\begin{figure}[t!]
\centerline{\includegraphics[height=5.cm]{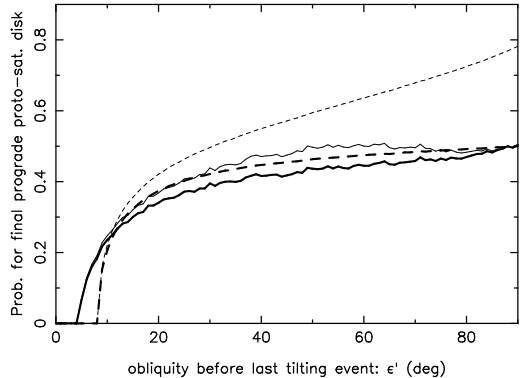}}
\caption{\small The probability that the proto-satellite disk has a
  prograde orientation relative to the final spin axis of Uranus as a
  function of the obliquity $\epsilon'$ that Uranus had prior to the
  final tilting. The results of {four} calculations are
  presented {according to the assumptions explained in the text.}
}
%For the dashed curves we assume that the
%    proto-satellite disk was equatorial before the final tilting; for
%    the solid curves we assumed that the disk precessed around the
%    equator with inclination $I=\epsilon'$; for the bold curves we
%    assume that all possible relative orientations of the original and
%    final spin vectors of the planet are equi-probable; for the thin
%    curve we assume that the probability to change the spin unitary
%    vector by $\vec{\delta\epsilon}$ is proportional to
%    $1/||\vec{\delta{\epsilon}}||$.}

\label{proba}
\end{figure}

We have computed, with Monte-Carlo simulations, the probability that
the final proto-satellite disk is prograde as a function of
$\epsilon'$. {We have done four simulations, adopting different
  assumptions; in each simulation $\epsilon'$ progresses from
  $0^\circ$ to $90^\circ$ by one-degree steps.}  First, we assume
that, when the planet had an obliquity $\epsilon'$, it was surrounded
by a proto-satellite disk that precessed rigidly around its equator
with an inclination $I=\epsilon'$. From the results in the previous
sections, this is expected if such obliquity had been acquired
impulsively from an initial obliquity of 0. The Monte Carlo
calculation in this case assumes two random angles: (i) the azimuthal
orientation of the final spin vector (of obliquity
$\epsilon=98^\circ$) relative to the plane of the Sun and (ii) the
precession phase of the proto-satellite disk at the time of the final
tilting. The result is illustrated with the bold solid curve in
Fig.~\ref{proba}. Second, we assume that, when the planet had an
obliquity $\epsilon'$, its proto-satellite disk had the time and the
conditions to become equatorial. In this case, the Monte Carlo
calculation has to assume just one random angle (the orientation of
the new vs. the old spin vectors). The result is illustrated with a
bold dashed curve in Fig.~\ref{proba}.  {Then, we relax the
  assumption that the relative orientations of the original and final
  spin vectors of the planet are equi-probable. Instead, we assume
  that probability that the spin vector has a change
  $\vec{\delta\epsilon}=\vec{\epsilon}-\vec{\epsilon'}$ is
  proportional to $1/||\vec{\delta\epsilon}||$ (here $\vec{\epsilon'}$
  and $\vec{\epsilon}$ are, respectively, the unitary spin vectors of
  the planet before and after the final tilting and $||.||$ denotes
  the Euclidean norm of a vector).  If the disk is precessing around
  the planetary equator before the final tilt, the result is
  illustrated by a thin solid curve and it is not very different from
  that obtained in the first calculation (compare with the bold solid
  curve). This is because the scalar product between the angular
  momentum vectors of the disk and the planet is more sensitive to the
  precession phase of the disk than to the relative orientation of the
  initial and final spin vectors of the planet. Instead, if the disk
  was equatorial, the result is illustrated by the thin dash curve and
  is significantly different from that obtained in the second
  calculation, for large values of $\epsilon'$ (compare with the the
  dash bold curve).}

 {As one can see in Fig. 3, the probability to have a prograde
   disk is null if $\epsilon'$ is smaller than 4 or 8 degrees for a
   rigid-precessing disk or an equatorial disk, respectively, for
   obvious geometric reasons. However,} in all cases we find that the
 probability { of ending up with a prograde satellite system}
 increases rapidly with $\epsilon'$ and, for $\epsilon '=30^\circ$
 (like the one of Neptune), it exceeds  40\%.

\section{Conclusions}

We conclude that the collisional tilting scenario for Uranus is
consistent with the prograde, equatorial character of the orbits of
its regular satellites, { as well as the size of the system}. The fact
that the satellites are prograde, implies that Uranus was not tilted
from 0 to 98 degrees in one shot. Instead, it had to have { had} a
non-negligible obliquity, { prior to} the { final} giant impact. Thus,
{ Uranus} should have experienced at least two giant collisions.

This result, together with the obliquity of Neptune, which also has no
other explanation than a collisional tilt, suggests that giant
impacts, affecting the obliquities, { were} rather common during the
growth of the ice-giants of the solar system. {Thus, these planets
  presumably did not grow by the sole accretion of small planetesimals
  as often envisioned.  Instead, towards the end of their accretion
  history, they should have experienced a phase similar to that
  characterising the process of terrestrial planet formation,
  i.e. dominated by giant impacts with other large planetary
  embryos. Past and future models of growth of giant planet cores
  should be confronted with this constraint.}

%\begin{thebibliography}{}

\section{References}

\begin{itemize}

\item[-]Bou{\'e}, G., 
Laskar, J.\ 2010.\ A Collisionless Scenario for Uranus Tilting.\ The 
Astrophysical Journal 712, L44-L47.

\item[-]
Canup, R.~M., Ward, 
W.~R.\ 2000.\ A Possible Impact Origin of the Uranian Satellite System.\ 
Bulletin of the American Astronomical Society 32, 1105. 

\item[-]Canup, R.~M., Ward, 
W.~R.\ 2002.\ Formation of the Galilean Satellites: Conditions of 
Accretion.\ The Astronomical Journal 124, 3404-3423. 

\item[-]
Charnoz, S., Salmon, 
J., Crida, A.\ 2010.\ The recent formation of Saturn's moonlets from 
viscous spreading of the main rings.\ Nature 465, 752-754. 

\item[-] Duncan, M.~J., Levison, 
H.~F., Lee, M.~H.\ 1998.\ A Multiple Time Step Symplectic Algorithm for 
Integrating Close Encounters.\ The Astronomical Journal 116, 2067-2077.

\item[-]Gaburov, E., Harfst, 
S., Portegies Zwart, S.\ 2009.\ SAPPORO: A way to turn your graphics cards 
into a GRAPE-6.\ New Astronomy 14, 630-637. 

\item[-]
Hahn, J.~M.\ 2003.\ The Secular Evolution of the Primordial Kuiper Belt.\ The 
Astrophysical Journal 595, 531-549. 

\item[-] Murray, C.D. and  Dermott, S.F. 1999. {\it Solar System
  Dynamics}. Cambridge University Press.

\item[-] Parisi, M.~G., Brunini, A.\ 1997.\ Constraints to Uranus' Great Collision-II.\ Planetary and Space Science 45, 181-187. 

\item[-]Safronov, V. S. 1966. Sizes of the largest bodies
falling  onto the planets during their formation. {\it Sov.
Astron.}, {9}, 987--991. 

\item[-]
Slattery, W.~L., Benz, 
W., Cameron, A.~G.~W.\ 1992.\ Giant impacts on a primitive Uranus.\ Icarus 
99, 167-174. 

\end{itemize}

%\end{thebibliography}
\end{document}